\documentclass[twocolumn,preprintnumbers,superscriptaddress,floatfix]{revtex4}
\usepackage{amsmath}
\usepackage{amssymb}
\usepackage{mathrsfs}
\usepackage{graphicx}
\usepackage{bm}
\usepackage{epstopdf}
\usepackage[colorlinks=true,citecolor=blue,linkcolor=blue,urlcolor=blue]{hyperref}
\usepackage{bookmark}
\usepackage{color}
\begin{document}
\textbf{\textcolor{red}{Copyright: This article has been submitted to the journal Frontiers of Physics}}
\title{
Electronic properties of 2{\it H}-stacking bilayer MoS$_2$ measured by terahertz time-domain spectroscopy}
\author{Xingjia Cheng}
\affiliation{Key Laboratory of Materials Physics, Institute of Solid State Physics, HFIPS, Chinese Academy of Sciences, Hefei 230031, China}
\affiliation{University of Science and Technology of China, Hefei 230026, China}

\author{Wen Xu}\email{wenxu$_$issp@aliyun.com}
\affiliation{Micro Optical Instruments Inc., Shenzhen 518118, China}
\affiliation{Key Laboratory of Materials Physics, Institute of Solid State Physics, HFIPS, Chinese Academy of Sciences, Hefei 230031, China}
\affiliation{School of Physics and Astronomy and  Yunnan Key Laboratory of Quantum Information of Yunnan Province, Yunnan University, Kunming 650091, China}

\author{Hua Wen}
\affiliation{Key Laboratory of Materials Physics, Institute of Solid State Physics, HFIPS, Chinese Academy of Sciences, Hefei 230031, China}
\affiliation{University of Science and Technology of China, Hefei 230026, China}

\author{Jing Zhang}
\affiliation{Key Laboratory of Materials Physics, Institute of Solid State Physics, HFIPS, Chinese Academy of Sciences, Hefei 230031, China}
\affiliation{University of Science and Technology of China, Hefei 230026, China}

\author{Heng Zhang}
\affiliation{Key Laboratory of Materials Physics, Institute of Solid State Physics, HFIPS, Chinese Academy of Sciences, Hefei 230031, China}
\affiliation{University of Science and Technology of China, Hefei 230026, China}

\author{Haowen Li}
\affiliation{Micro Optical Instruments Inc., 518118 Shenzhen, China}

\author{Qingqing Chen}
\affiliation{Key Laboratory of Materials Physics, Institute of Solid State Physics, HFIPS, Chinese Academy of Sciences, Hefei 230031, China}
\affiliation{University of Science and Technology of China, Hefei 230026, China}

\date{\today}


\begin{abstract}
Bilayer (BL) molybdenum disulfide (MoS$_2$) is one of the most important electronic structures not only in valleytronics but also in realizing twistronic systems on the basis of the topological mosaics in Moir\'e superlattices. In this work, BL MoS$_2$ on sapphire substrate with 2$H$-stacking structure is fabricated. We apply the terahertz (THz) time-domain spectroscopy (TDS) for examining the basic optoelectronic properties of this kind of BL MoS$_2$. The optical conductivity of BL MoS$_2$ is obtained in temperature regime from 80 to 280 K. Through fitting the experimental data with the theoretical formula, the key sample parameters of BL MoS$_2$ can be determined, such as the electron density, the electronic relaxation time and the electronic localization factor. The temperature dependence of these parameters is examined and analyzed. We find that, similar to monolayer (ML) MoS$_2$, BL MoS$_2$ with 2$H$-stacking can respond strongly to THz radiation field and show semiconductor-like optoelectronic features. The theoretical calculations using density functional theory (DFT) can help us to further understand why the THz optoelectronic properties of BL MoS$_2$ differ from those observed for ML MoS$_2$. The results obtained from this study indicate that the THz TDS can be applied suitably to study the optoelectronic properties of BL MoS$_2$ based twistronic systems for novel applications as optical and optoelectronic materials and devices.
\end{abstract}

\maketitle
\setlength{\parskip}{0.1em}
\section{Introduction}
Since the discovery of graphene in 2004 \cite{1}, the investigation of atomically thin or two-dimensional (2D) electronic systems \cite{2} has quickly become one of the hot and fast-growing fields of research in condensed matter physics, material science, electronics and optoelectronics. At present, one of the most popularly studied 2D electronic systems is the few-layer and even monolayer (ML) transition metal dichalcogenides (TMDs) such as molybdenum disulfide (MoS$_2$) \cite{3} and tungsten disulfide (WS$_2$) \cite{4}. It has been found that the ML TMDs can exhibit the unique valleytronic properties which can be applied for, e.g., information storage and processing \cite{5}. More interestingly, bilayer (BL) TMDs can be utilized to form the van der Waals (vdW) heterostructures in which the topological mosaics in Moir$\acute{\rm e}$ superlattices can be presented \cite{6}. In such a case, the bilayer structures can offer an added layer degree of freedom. The interactions between two TMD layers with specific stacking order can significantly modify the electronic, optical \cite{6} and vibrational properties of the electronic system. Thus, the twistronic features of the BL TMD system can be observed by tuning the vdW heterojunction from normal to the inverted type-II regime through, e.g., taking different stacking structures and/or applying an inter-layer bias voltage \cite{6}. It is known that, similar to high-T$_c$ superconductors, superconductivity can be observed in BL graphene based twistronic systems \cite{7}. As for BL TMD based twistronic structures, novel electronic and optoelectronic phenomena, such as infrared inter-layer Moir\'e excitons \cite{8}, Dirac fermions with massive Dirac cones \cite{9}, on/off switching of the topological band inversion \cite{6}, to mention but a few, have been observed and studied. Therefore, it is important and significant to study the basic optoelectronic properties of the BL TMD systems, which becomes the prime motivation of the present research work.

In the present study, we focus our attention on one of the most popularly investigated BL TMD materials such as BL MoS$_2$. MoS$_2$ is a hexagonal crystal in which molybdenum is sandwiched between the sulfur atoms on both sides. Therefore, BL MoS$_2$ can have six types of high symmetrical stacking structures \cite{6} noted as $H_X^X$, $H_M^M$, $H_X^M$, $R_X^M$, $R_M^X$ and $R_M^M$ depending on the symmetry of the stacking configurations. Among them, $H_X^M$ (or 2$H$) is commonly formed BL stacking structure with a relatively high stacking symmetry. The other stacking configurations can be achieved by twisting and/or translating one of the MoS$_2$ layer away from 2$H$-stacking. In contrast to gapless graphene, ML, BL and bulk MoS$_2$ can have energy gaps between the conduction and valence bands, depending on the numbers and the stacking structures of the MoS$_2$ layers \cite{10,11}. The band gap $E_g$ around the $K$-point decreases with increasing the number of the MoS$_2$ layers. It is found that $E_g\simeq$ 1.8 eV, 1.6 eV and 1.29 eV for respectively ML, BL and bulk MoS$_2$ \cite{10,11,12}. Moreover, it is known that ML MoS$_2$ is with a direct band gap around the $K$- and $K'$-point \cite{10}. In contrast, BL MoS$_2$ is with a weak direct band gap or even indirect band gap around the $K$-point, depending also on the stacking order. For 2$H$-stacking BL MoS$_2$, the weak direct band gap can be presented around the $K$-point and the photoluminescence (PL) can be measured experimentally, although the PL efficiency of the BL MoS$_2$ is much lower in comparison to that of the  ML MoS$_2$ \cite{10}. In addition, both ML and BL of MoS$_2$ show different degrees of band splitting at the top of the valence band around the $K$-point. For BL MoS$_2$, the effect of the band splitting is particularly pronounced for 2$H$- and 3$R$ (or $R_X^M$)-stacking configurations \cite{13}. These features of the electronic band structures in ML, BL and bulk MoS$_2$ have been confirmed experimentally by the results obtained from, e.g., PL measurements \cite{10,14}. At present, the optical and optoelectronic properties of BL MoS$_2$ have been mainly studied in visible \cite{15} and infrared \cite{14} bandwidths, where the photon energy $\hbar\omega$ is larger than $E_g$, to look into mainly the effects of intra- and inter-layer excitons \cite{6,8}. Very recently, terahertz (THz) optoelectronic techniques have been applied for the investigation of BL MoS$_2$.  The modulation of the optical conductivity of BL TMD heterojunctions was studied by varying the excitation optical power using light-pumped THz time-domain measurements \cite{49}. The photovoltaic properties of BL TMD heterojunctions was also studied by examining the delay between optical pumping and THz measurements \cite{50}. We note that these investigations are based on the optical pump and THz probe (OPTP) technique in which the photon energy of pump radiation is larger than the energy gap of BL MoS$_2$. Therefore, these experimental work examined the optical response induced mainly by inter-band electronic transition accompanied by photon-generated carriers and associated exciton effect \cite{49,50}. The electronic relaxation time measured via OPTP normally corresponds to inter-band energy relaxation time.

In this work, we intend studying THz optoelectronic properties of BL MoS$_2$. Because THz photon energy is much smaller than the band gap of BL MoS$_2$ ($f=\omega/2\pi$ = 1 THz is about 4.13 meV), THz radiation normally does not cause the photon-induced carriers and associated intra- and inter-layer excitons in BL MoS$_2$. Thus, we can study the response of free-carriers in BL MoS$_2$ to the applied THz radiation field. In recent years, we have applied the technique of THz time-domain spectroscopy (TDS) for the investigation of optoelectronic \cite{16} and magneto-optical \cite{17} properties of ML MoS$_2$. In the present study, we would like to apply the similar technique to examine the electronic dynamics in BL MoS$_2$. Furthermore, because 2$H$-stacking is the most basic BL structure for MoS$_2$ and for other BL TMD systems, here we would like to contribute a systematic experimental investigation into THz optoelectronic properties of 2$H$-stacking BL  MoS$_2$. Our aim of this study is at gaining an in-depth understanding of the basic physical properties of BL TMD materials for potential applications as advanced electronic and optoelectronic devices.

\section{Experimental and theoretical studies}
\subsection{Sample fabrication}

\begin{figure*}
\centering
\includegraphics[width=15.6cm]{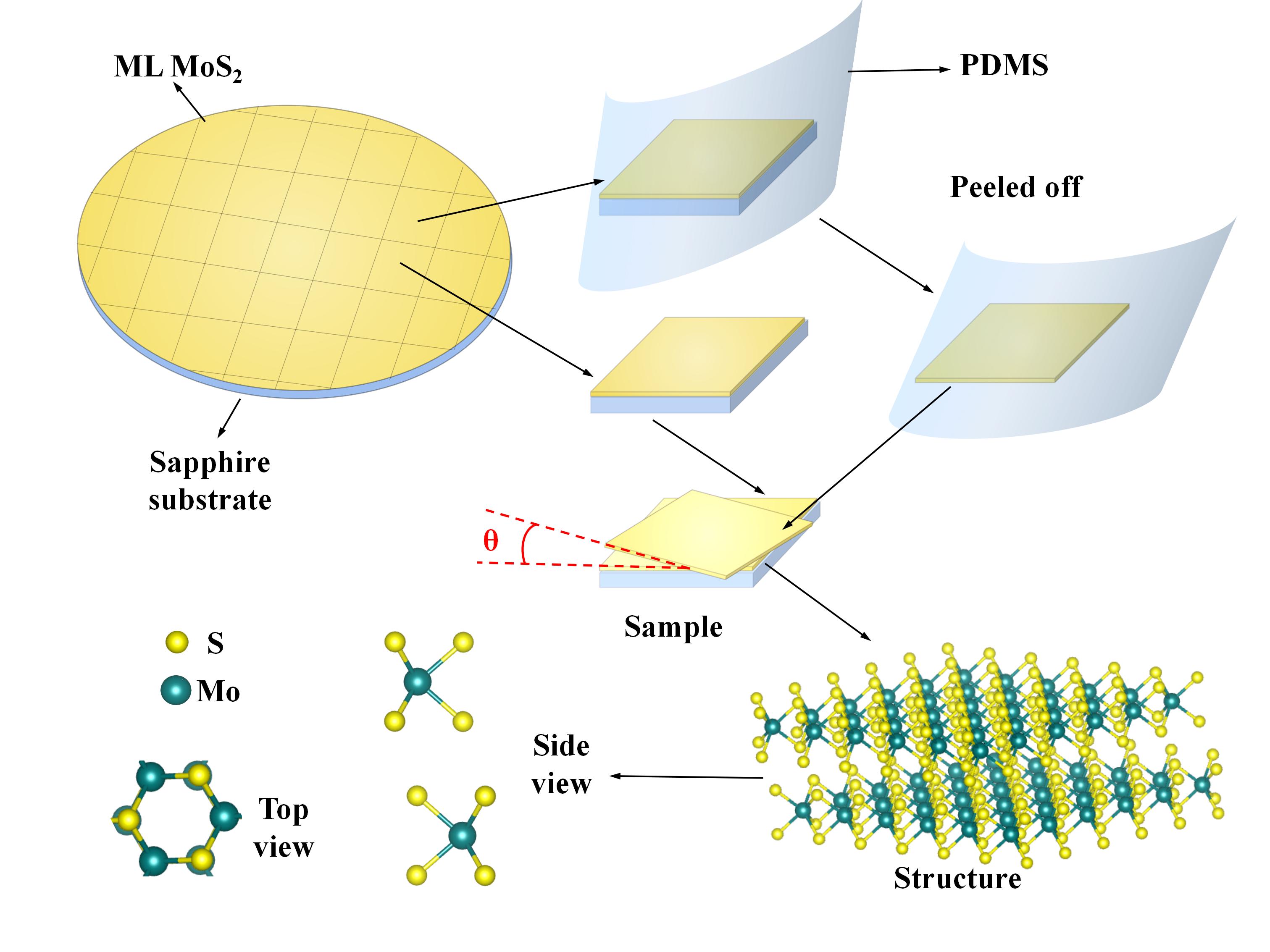}
\caption{Schematic diagram of the sample preparation. i) The ML MoS$_2$ was grown on sapphire wafer by using the CVD; ii) The as-grown ML-MoS$_2$/sapphire wafer was cut into small pieces; iii) The ML MoS$_2$ film on a small piece was peeled off from sapphire substrate; And iv) the ML MoS$_2$ film was
transferred onto the small ML-MoS$_2$/sapphire piece, where the twisting angle between two ML MoS$_2$ films was set to be at $\theta=60^\circ$ to obtain the 2$\it{H}$-stacking BL MoS$_2$. The bottom figures show lattice structure of 2$\it{H}$-stacking BL MoS$_2$.}
\label{fig1}
\end{figure*}\normalsize

The schematic diagram of the sample preparation in this study is shown in Fig. \ref{fig1}. i) The BL MoS$_2$ on sapphire substrate was fabricated by using the standard technique of the chemical vapor deposition (CVD) \cite{20,21,22,23}. MoO$_3$ (99.999\% purity) and solid sulfur (99.999\% purity) were applied as the molybdenum source and sulfur source, respectively, and argon was taken as the growth carrier gas. The sample growth was carried out in a dual temperature zone tube furnace with a diameter of 80 mm. The growth conditions were as follows: MoO$_3$ was heated to 650 $^{\circ}$C, sulfur was heated to 180 $^{\circ}$C, the growth pressure was at 4000 Pa, and the reaction time was 10 minutes. By taking these growth conditions, the ML MoS$_2$ on sapphire wafer can be fabricated. ii) The slicing machine was used to cut the ML-MoS$_2$/sapphire-wafer into multiple 1$\times$1 cm$^2$ square pieces. The orientation of ML MoS$_2$ film on these pieces is the same \cite{18}. When taking each small piece of the sample, we ensured that the original direction of the samples was the same. iii) The ML MoS$_2$ film on a small piece was peeled off from sapphire wafer with the help of polydimethylsiloxane (PDMS) films. The deionized water penetrated through the edges of the ML MoS$_2$ and the sapphire wafer to the interface between the ML MoS$_2$ and the sapphire. Thus, we are able to separate the MoS$_2$ film from the sapphire wafer so that we could slowly get the PDMS/MoS$_2$ film from the sapphire wafer. iv) The PDMS/ML MoS$_2$ film was then transferred onto the surface of ML-MoS$_2$/sapphire piece. The twisting angle between two ML MoS$_2$ films should be set to be at $\theta=60^\circ$ in order to obtain the 2$\it{H}$-stacking BL MoS$_2$. We first aligned two MoS$_2$ films on PDMS and on sapphire, then used the rotating stage to rotate the ML MoS$_2$ film on PDMS to 60$^{\circ}$ angle relative to the ML MoS$_2$ film on sapphire. The PDMS film was then mechanically peeled off and, finally, the 2$\it{H}$-stacking BL MoS$_2$ on sapphire substrate was prepared. This technique of sample growth and preparation is similar to those reported by other research groups \cite{18,19}. It should be noted that due to the presence of the vdW force between two MoS$_2$ layers, 2$H$-stacking type of the BL MoS$_2$ can normally obtained by using this technique.

The size of the BL MoS$_2$/sapphire sample used in this study is 1$\times$1 cm$^2$ and the thickness of the sapphire substrate is 0.3 mm. We note that the diameter of the THz light beam focused on the sample is about 3 $\sim$ 5 mm in our experimental measurement system. Thus, the BL MoS$_2$ sample prepared in this study is big enough for the THz TDS measurement. It should be noted that the preparation of other types of BL MoS$_2$ structures, e.g., 3$\it{R}$-stacking, requires the relative translation between two MoS$_2$ layers, which is on nanometer distance scale. This requires considerable research work and we do not attempt it in the present study.

\subsection{Sample characterization}
\begin{figure*}
\centering
\includegraphics[width=15.6cm]{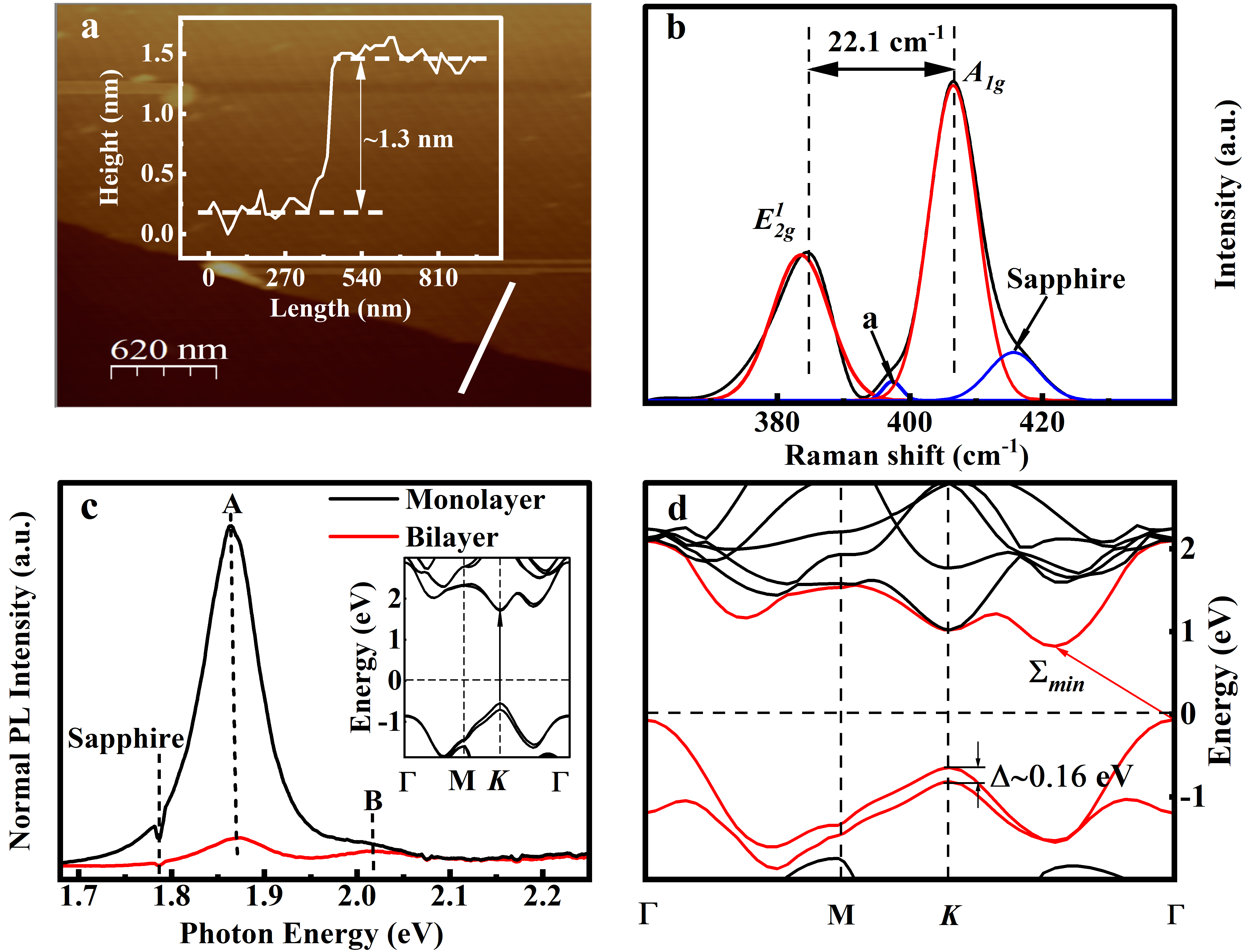}
\caption{(a) AFM image of BL MoS$_2$ on sapphire substrate. The inset shows the hight of the step-change between MoS$_2$ layer and the substrate along the line marked as white straight line, which corresponds to the thickness of the BL MoS$_2$. (b) Raman spectra of BL MoS$_2$ on sapphire substrate (black curve), measured at room-temperature by a 532 nm laser excitation. Two characteristic peaks for BL MoS$_2$, $A_{1g}$ and $E^1_{2g}$ with a spacing about 22.1 cm$^{-1}$, can be clearly identified. By spectral decomposition (red and blue curves), the Raman peaks $A_{1g}$ and $E^1_{2g}$ along with ``a'' peak induced characteristically by the 2$H$-stacking type and the characteristic peak for sapphire can be found. (c) Photoluminescence (PL) spectra of ML (black curve) and BL (red curve) MoS$_2$, measured at room-temperature using a 532 nm laser excitation, where the results are shown after deducting the signals from sapphire substrate. The inset shows the electronic band structure of ML MoS$_2$ obtained from DFT calculation. (d) Electronic band structure of 2$H$-stacking BL MoS$_2$, obtained from the DFT calculation. Here the $\Sigma_{min}$-point in the conduction band is lower than that in the $K$-point, with a difference about 197 meV.}
\label{fig2}
\end{figure*}\normalsize

In this study, the basic features of BL MoS$_2$ on sapphire were characterized experimentally by using atomic force microscope (AFM), Raman spectroscopy, and photoluminescence(PL). Fig. \ref{fig2}(a) shows the results obtained from AFM on the sample. From AFM imaging, we can see that the MoS$_2$ film (lighter area) on sapphire substrate (darker area) is clean and smooth. By measuring the hight of the step-change between MoS$_2$ layer and the substrate along the line marked as white straight line in Fig. \ref{fig2}(a), we can determine the thickness of the MoS$_2$ film to be about 1.3 nm. Because the thickness of ML MoS$_2$ is about 0.65 nm \cite{21}, the step-change of 1.3 nm in Fig. \ref{fig2}(a) suggests that the film is a BL MoS$_2$ structure \cite{21,22}.

Fig. \ref{fig2}(b) shows the Raman spectrum of BL MoS$_2$ on sapphire substrate using a 532 nm laser excitation at room-temperature. Two characteristic Raman peaks can be measured, which correspond respectively to in-plane Raman mode $E^1_{2g}$ and out-of-plane Raman model $A_{1g}$\cite{22}. The wavenumber difference of the two peaks, $A_{1g}$-$E^1_{2g}$, is about 22.1 cm$^{-1}$, in consistent with the result obtained for BL MoS$_2$ \cite{21,22}. Moreover, by carefully analyzing the Raman spectrum via two Gaussian fittings (red and blue curves), a peak ``a'' at about 396 cm$^{-1}$ can be found, which corresponds to 2$H$-stacking BL MoS$_2$ \cite{14}. The peak at $\sim 417$ cm$^{-1}$ is the characteristic Raman peak of the sapphire substrate \cite{22}.

We apply a He-Ne laser (532 nm) as the light excitation source for the PL measurement at room temperature, with a grating of 1800 I/mm and the exposure time of 10 s. The signal reception is charge-coupled device (CCD) detector. The PL spectra for ML and BL MoS$_2$ on sapphire substrate and for sapphire substrate alone are measured respectively. The results shown in Fig. \ref{fig2}(c) are obtained by deducting the PL spectrum of the substrate from that for ML or BL MoS$_2$/substrate. In doing so, we are able to see more clearly those two PL peaks for 2$H$-stacking BL MoS$_2$.
The PL spectrum of BL MoS$_2$ on sapphire substrate (red curve) is shown in Fig. \ref{fig2}(c). For comparison, we also present the result for ML MoS$_2$ on sapphire substrate (black curve) prepared by using the similar CVD technique \cite{21,22,23}.
Two PL peaks, A at $\sim$1.86 eV and B at $\sim$2.01 eV, can be observed. These two peaks are induced by excitonic transitions from split valence bands, with an energy spacing $\Delta\sim$ 0.15 eV measured experimentally from energy distance between A and B peaks in Fig. 2(c) and $\Delta\sim$ 0.16 eV obtained by our DFT calculations, to the conduction band around the $K$-point in ML and BL MoS$_2$ (see Fig. \ref{fig2}(d)). These results agree with those reported previously by other groups \cite{15,24}. As can be expected, the intensity of PL emission from BL MoS$_2$ is significantly weaker than that from ML MoS$_2$. These results can be further understood with the help of the electronic band structure (see Fig. \ref{fig2}(d)) of 2$H$-stacking BL MoS$_2$ structure (Fig. \ref{fig1}), obtained from first principle calculation on the basis of DFT \cite{25}.

From above mentioned experimental measurements and findings, we can confirm that the sample used in this study is BL MoS$_2$ with 2$H$-stacking configuration. Furthermore, the BL MoS$_2$ sample used in this study is n-type, confirmed by our Hall measurement. We note that it is not easy to fabricate the Hall bar ohmic contact electrodes on BL MoS$_2$ sample. We only used the Hall measurement to identify the n-type feature of the sample. It has been shown that BL MoS$_2$ with other stacking orders can also be grown by using the CVD technique and be realized by taking the PDMS approach \cite{21,22,23,24}.

\subsection{Setup of THz time-domain spectroscopy}

\begin{figure}
\centering
\includegraphics[width=8.6cm]{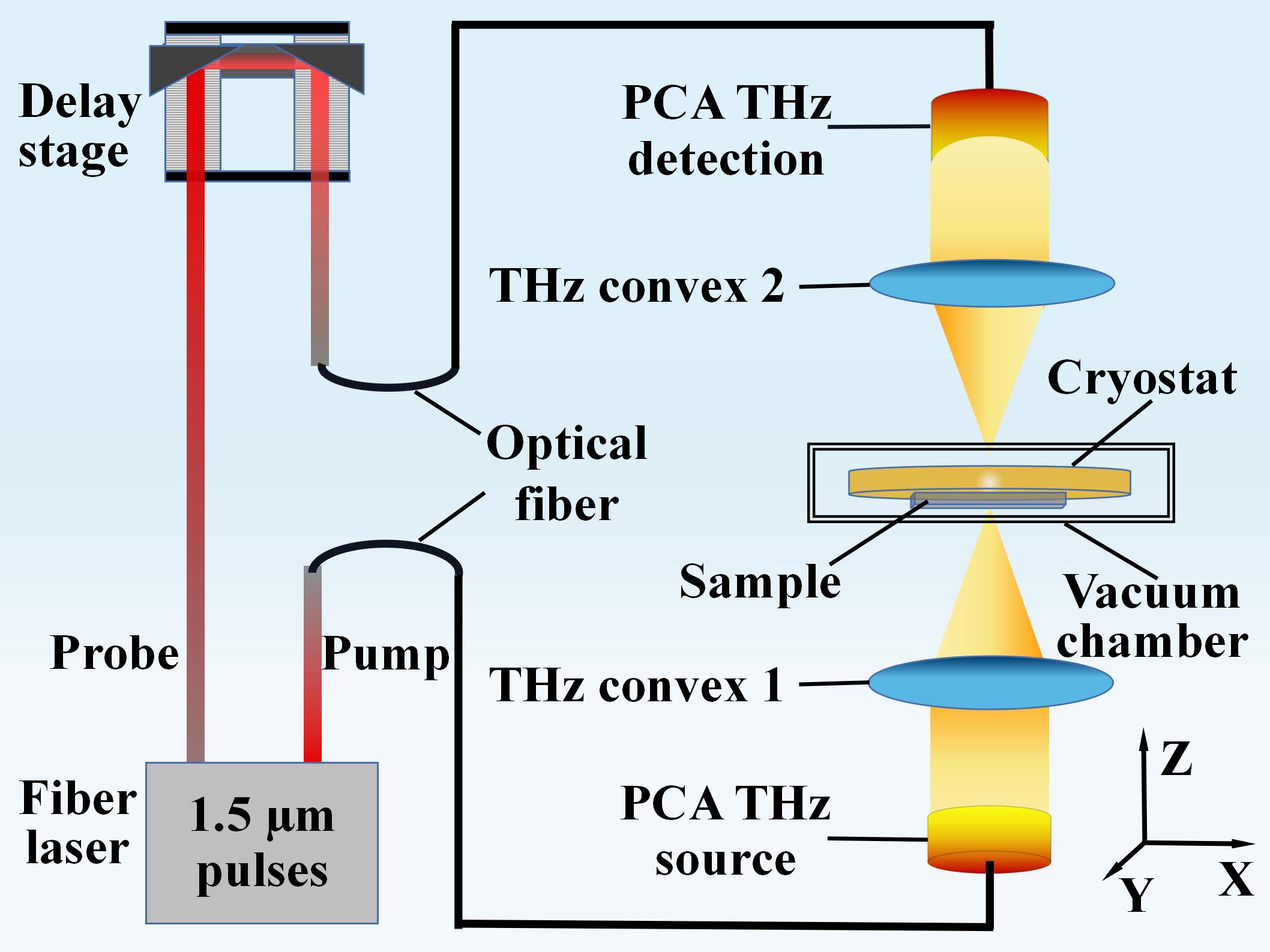}
\caption{Schematic diagram of the THz TDS system for optical transmission measurement. Here PCA is the photoconductance antenna.}
\label{fig3}
\vspace{-0.5cm}
\end{figure}\normalsize

The schematic diagram of the THz TDS system used in this study is shown in Fig. \ref{fig3}. i) The pumping light source is a femtosecond (fs) fiber laser (ROI optoelectronics) with 1550 nm wavelength, 80 fs pulse width and 100 MHz repetition frequency. ii) The fs laser is divided into two beams. The higher intensity beam is the pump light, which irradiates on an InGaAs photoconductance antenna (PCA. Melno, Germany) to generate the broadband THz pulses. The lower intensity fs laser beam irradiates, with a time delay, on another InGaAs PCA for the detection of THz transmission through the sample via photo-electric sampling. iii) The generated THz beam is a horizontally polarized field, which is directed towards the sample surface normally. And iv) the sample is fixed on a sample holder in the cryostat (ST-500, Jains) with a quartz window, and the sample chamber is in a vacuum. Since the cryostat has a quartz window and the quartz can only transmit efficiently the THz light beam up to 1.2 THz \cite{27,28}, in the measurement we used the effective THz frequency ranging from 0.2 THz to 1.2 THz. In the present study, the variation of the temperature was down to liquid nitrogen temperature at about 80 K. Thus, we were able to measure the strength of the THz electric field transmitted through the sample as a function of the delay time in a temperature range from 80 K to 280 K. Moreover, the experimental system was slowly flushed with dry nitrogen gas to prevent the absorption of the THz waves by moisture and dusts in free-space.

\subsection{Theoretical calculation}
By taking the lattice structure of 2$H$-stacking BL MoS$_2$ as shown in Fig. \ref{fig1}, we study the corresponding electronic band structure via the DFT calculation as implemented in the VASP code using projector augmented wave pseudo-potentials and plane-wave expansions with a cut-off energy of 500 eV. In the calculation, the lattice parameter $a=3.161$ \AA, the Mo-Mo inter-layer distance $d=6.147$ \AA, and the Mo-S interatomic distance $l=1.584$ {\AA} along the [0001] direction of a relaxed 2$H$-stacking BL MoS$_2$ crystal were taken in line with the corresponding experimental values \cite{26}. A $K$-point mesh of $15\times 15\times 1$ in the first Brillouin zone was found to yield well-converged results. A vacuum space thickness of 20 {\AA}  is used to prevent the interactions between the adjacent periodic images of the BL structure. The atomic positions were optimized until all components of the forces on each atom were reduced to the values below 0.001 eV/\AA. The exchange-correlation function was treated within the Perdew-Burke-Ernzerhof (PBE) generalized gradient approximations (GGA). In the calculation, we consider the vdW force between two MoS$_2$ layers, setting the IVDW value to be 11. The electronic energy spectrum of 2$H$-stacking BL MoS$_2$ obtained from this calculation is shown in Fig. \ref{fig2}(d). We also calculated the electronic band structure of ML layer for comparison (see inset in Fig. 2(c)). Here we note that the calculation of the electronic band structure in ML MoS$_2$ used the hybrid density functional, whereas that in BL MoS$_2$  used the PBE functional. The effect of interlayer vdW forces was taken into account when calculating the electronic band structure of BL MoS$_2$. Otherwise the conditions for both calculations are basically the same. The detailed discussions of the electronic band structure of 2$H$-stacking BL MoS$_2$ are presented in the section of Discussions.

\section{Results and discussions}
\subsection{THz transmission spectrum and optical conductivity}
\begin{figure}
\includegraphics[width=8.6cm]{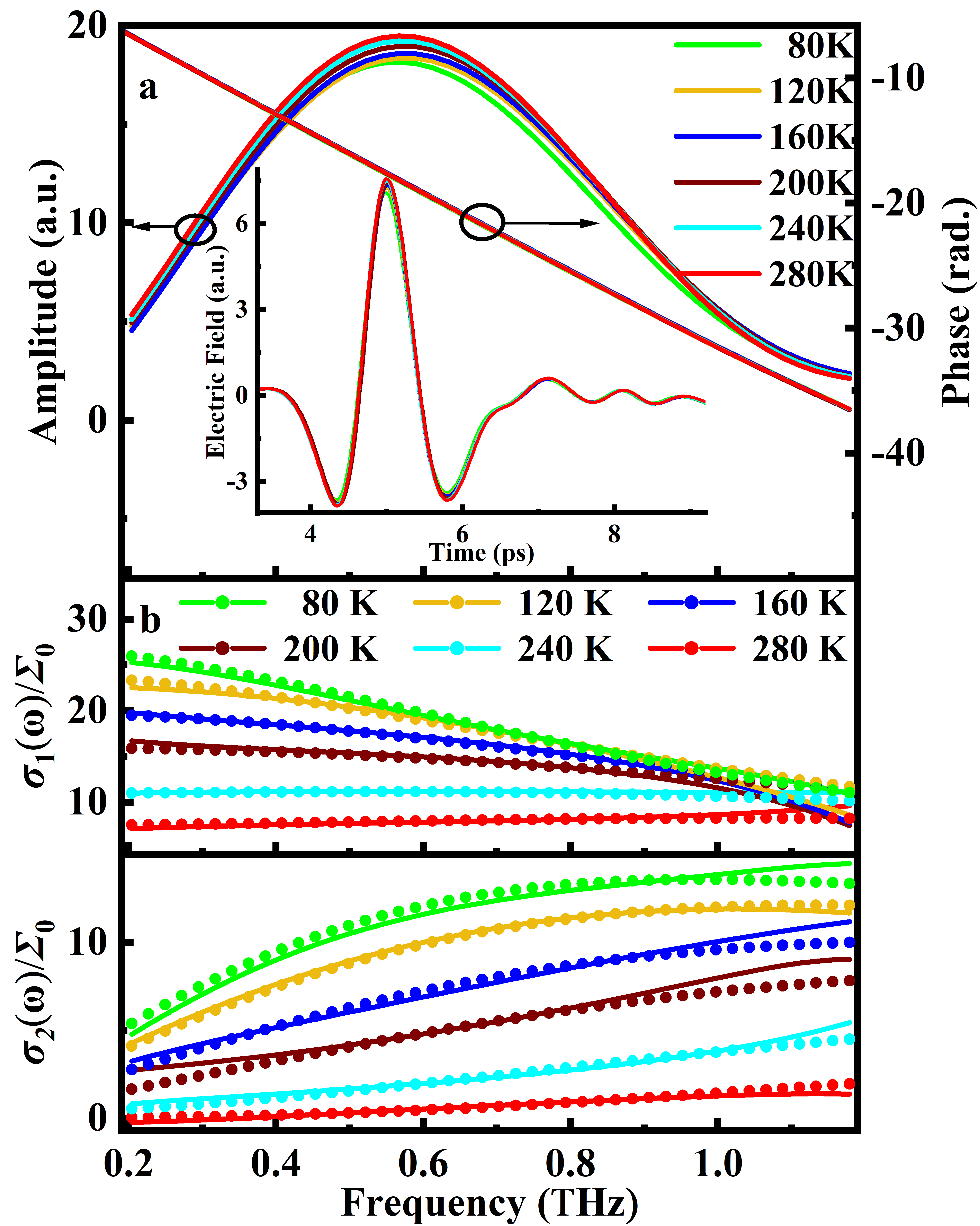}
\caption{(a) Amplitude and phase angle of the THz electric field $E_{ms}(\omega)$ transmitted through a BL MoS$_2$-sapphire sample as a function of radiation frequency $f=\omega/2\pi$ at different temperatures. The inset shows the electric field strength $E_{ms}(t)$ of the THz beam transmitted through a BL MoS$_2$/sapphire sample as a function of delay time at different temperatures. (b) Real $\sigma_1(\omega)$ (upper panel) and imaginary $\sigma_2(\omega)$ (lower panel) parts of the optical conductivity for BL MoS$_2$ as a function of radiation frequency f = $\omega/2\pi$ at different temperatures. The dots are experimental results and the curves are obtained from Drude-Smith formula given by Eq. (\ref{1}). Here $\Sigma_0 = e^2/4\hbar = 6.07\times10^{-5}$ S.}
\label{fig4}
\end{figure}

Applying transmission experiment based on THz TDS, we can measure the electric field strength transmitted through sapphire substrate $E_s(t)$ and through the sample (i.e., BL MoS$_2$ on substrate) $E_{ms}(t)$ in time-domain. In Fig. \ref{fig4}(a), we show $E_{ms}(t)$ (inset) measured at different temperatures. As can been seen (see inset in Fig. \ref{fig4}(a)), $E_{ms}(t)$ depends sensitively on temperature. By Fourier transformation of $E_s(t)$ and $E_{ms}(t)$, we can obtain the corresponding electric field strength transmitted through sapphire substrate $E_s(\omega)$ and through the sample $E_{ms}(\omega)$ in frequency-domain, where $E_j(\omega)=|E_j(\omega)|e^{i\phi_j (\omega)}$. In Fig. \ref{fig4}(a), we show the modulus $|E_{ms}(\omega)|$ and the phase angle $\phi_{ms} (\omega)$ as a function of radiation frequency $f=\omega/2\pi$ for different temperatures. It is interesting to note that $|E_{ms}(\omega)|$ depends strongly on temperature, whereas $\phi_{ms} (\omega)$ shows a weak dependence upon temperature for BL MoS$_2$ on sapphire substrate.

\normalsize

From $E_s(\omega)$ and $E_{ms}(\omega)$, we can approximately evaluate the transmission coefficient for BL MoS$_2$ via \cite{29}: $t_m(\omega)\simeq E_{ms}(\omega)/E_{s}(\omega)$, where the effect of the optical reflections from MoS$_2$ surface and from BL MoS$_2$/substrate interface are neglected \cite{29}. It is known that optical conductivity, $\sigma(\omega)$, is a key and central physics quantity from which the optical coefficients such as absorption, transmission and reflection can be obtained by using the basic physics laws \cite{30}. It is also a bridge quantity connecting optics and condensed matter physics. From $t_m(\omega)$ we can determine $\sigma(\omega)$ for BL MoS$_2$ via Tinkham formula \cite{31}
\begin{equation}
t_m(\omega)={1+N_s \over 1+N_s +Z_0 \sigma(\omega)},
\end{equation}
where $N_s$ is the index of refraction for the substrate, $N_s=3.07$ for sapphire \cite{21,22}, and $Z_0 \approx 377 \ \Omega$ is the impedance of the free space. Because $t_m(\omega)$ is a complex quantity obtained via Fourier transformation, the optical conductivity is also a complex quantity $\sigma(\omega)=\sigma_1(\omega)+i \sigma_2(\omega)$. It is known that $\sigma_1(\omega)$ corresponds to the energy-consuming process or optical absorption, whereas $\sigma_2(\omega)$ describes the energy exchanging process between electrons and the radiation field. The real and imaginary parts of the optical conductivity for BL MoS$_2$ on sapphire are shown in Fig. \ref{fig4}(b) as a function of radiation frequency for different temperatures. The results obtained experimentally are shown by dotted curves. We find that with increasing radiation frequency $f=\omega/2\pi$, $\sigma_1(\omega)$ decreases in low-temperature regime $80$ K $\leq T \leq 200$ K and increases at relatively high-temperatures $T$ = 240 K and 280 K. $\sigma_2(\omega)$ increases with $\omega$ in temperature regime from $80$ K to $280$ K. Both $\sigma_1(\omega)$ and $\sigma_2(\omega)$ increases with decreasing temperature.

\subsection{Key sample parameters}

\begin{figure}
\includegraphics[width=8.6cm]{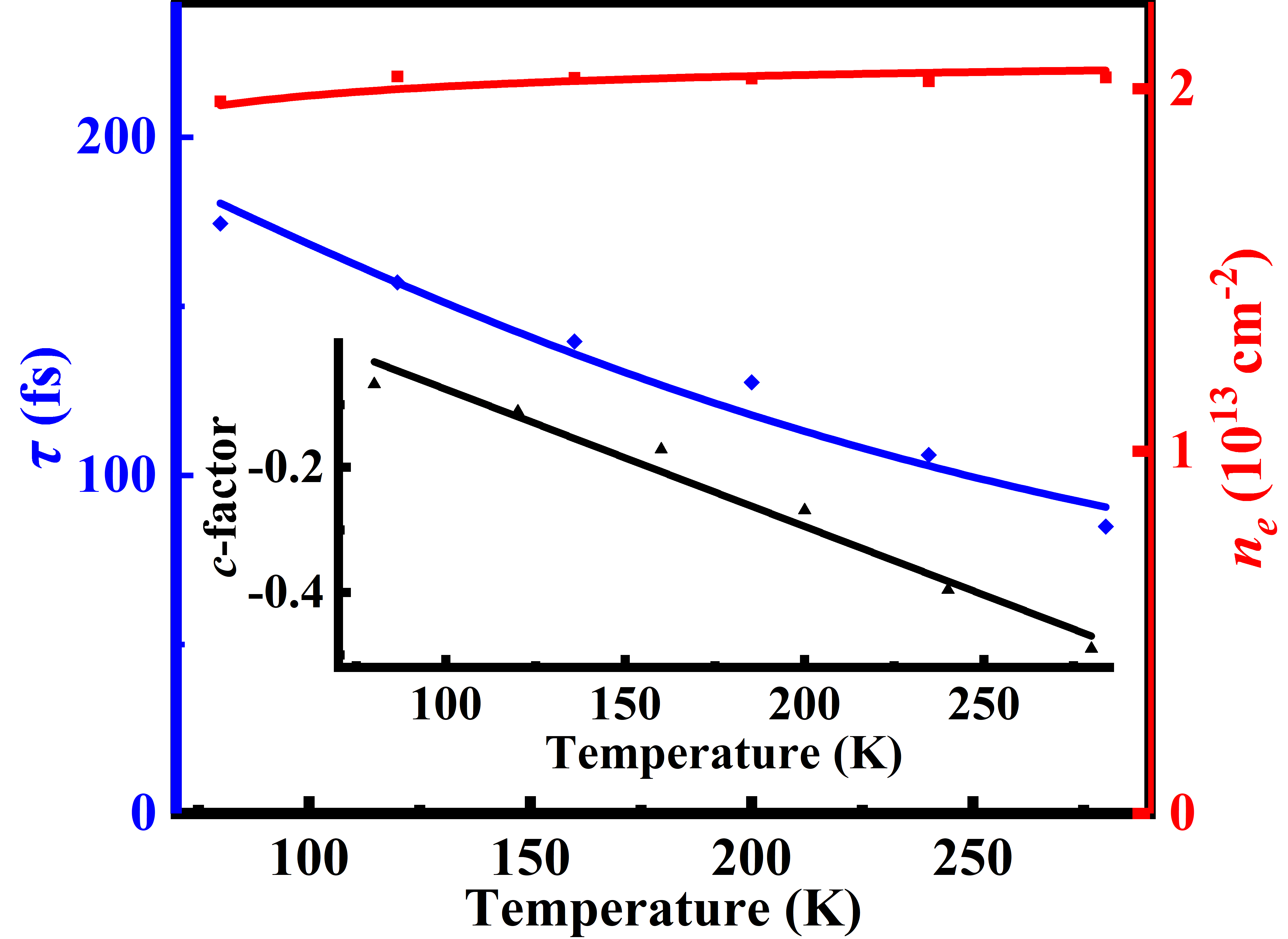}
\caption{Electron density n$_e$ (blue), electronic relaxation time $\tau$ (red), and electronic localization factor $c$ (inset) in 2$H$-stacking BL MoS$_2$ as a function of temperature. For n$_e$ and $c$, the curves are drawn to guide the eye, and for $\tau$ the curve is obtained by using Eq. (\ref{2}).}
\label{fig5}
\end{figure}

From a view point of condensed matter physics and electronics, optical conductivity $\sigma(\omega)$ in an electronic material is mainly determined by electronic band structure, electronic scattering mechanisms such as impurities and phonons and by electron interactions with incident photons \cite{32}. Because THz radiation does not cause photon-induced carriers in the electronic system, THz TDS measures mainly the electronic dynamics induced by free-carrier response to the radiation field. For a n-type BL MoS$_2$ sample, the optical conductivity is mainly determined by electronic transition within the conduction band. It is known that the commonly used theoretical approach to describe the optical conductivity for free-electrons is the Drude formula given as \cite{33}
\begin{equation}
\sigma(\omega)={\sigma_0 \over 1-i\omega\tau}={\sigma_0 \over 1+(\omega\tau)^2}+i {\sigma_0\omega\tau \over 1+(\omega\tau)^2},
\end{equation}
where $\sigma_0=e^2n_e\tau/m^*$ is direct current (DC) conductivity, $n_e$ is the electron density in the sample, $\tau$ is the electronic momentum relaxation time, and $m^*$ is the effective electron mass. The Drude formula suggests that $\sigma_1(\omega)$ should decrease with increasing $\omega$ and $\sigma_2(\omega)$ should first increase then decrease with increasing $\omega$. As shown in Fig. \ref{fig4}(b), both $\sigma_1(\omega)$ and $\sigma_2(\omega)$ obtained experimentally for BL MoS$_2$ on sapphire substrate do not obey the conventional Drude formula. In this study, we employ the Drude-Smith formula \cite{34} for the understanding of our experimental results. By taking only the first collision term in the general Drude-Smith formula \cite{34,35} into consideration, the Drude-Smith formula can be written as
\begin{equation}
\sigma(\omega)={\sigma_0 \over 1-i\omega\tau}(1+{c\over1-i\omega\tau}),\label{1}
\end{equation}
where $c=[-1,0]$ refers to the fraction of original velocity for an electron after a collision event, which corresponds to the effect of photon-induced electronic backscattering or localization \cite{34}. We find that both $\sigma_1(\omega)$ and $\sigma_2(\omega)$ given by Drude-Smith formula can fit very well to those obtained experimentally, as shown in Fig. \ref{fig4}(b) (see solid curves). Through fitting the experimental data with the theoretical formula, we can obtain the DC conductivity $\sigma_0$, the electronic relaxation time $\tau$ and the electronic localization factor $c$ for BL MoS$_2$. Moreover, it should be noted that in contrast to ML MoS$_2$ in which the conduction band minima is at the $K$-point, the bottom of the conduction band in 2$H$-stacking BL MoS$_2$ is at the $\Sigma_{min}$-point (see Fig. \ref{fig2}(d)). By taking the effective electron mass for 2$H$-stacking BL MoS$_2$ as \cite{36} $m^*\simeq 0.55 m_e$, with $m_e$ being the rest electron mass, we can determine the electron density $n_e$ in BL MoS$_2$ through the results obtained experimentally for $\sigma_0$. In Fig. \ref{fig5} we show the electron density $n_e$ (the blue section), the electronic relaxation time $\tau$ (the red section) and the electronic localization factor $c$ (inset) as a function of temperature for BL MoS$_2$. As can be seen, i) $n_e$ depends rather weakly on temperature in the regime from 80 K to 280 K, in line with experimental results reported by other research groups \cite{37}; ii) $\tau$ decreases with increasing temperature, which is a typical feature of a semiconductor \cite{31}; and iii) $|c|$ increases with temperature, implying that the photon-induced electronic backscattering or localization in 2$H$-stacking BL MoS$_2$ increases with temperature.
\normalsize
Since BL MoS$_2$ is a semiconductor-like material, the electron-impurity and electron-phonon interactions are major channels for electronic scattering in the material. In the present study, we take a simple formula:
\begin{equation}
{1\over\tau}={1\over\tau_I}+\alpha_{AC}T+N_0 \Gamma_{LO},\label{2}
\end{equation}
to fit the temperature dependence of the electronic relaxation time \cite{38,39} for BL MoS$_2$. Here, i) the first term $1/\tau_I$ is induced by electron-impurity scattering, which depends weakly on temperature \cite{39}; ii) the second term $\alpha_{AC}T$ comes from electron interaction with acoustic-phonons, which depends normally linearly on temperature \cite{38}; iii) the third term $N_0 \Gamma_{LO}$ is attributed from electron coupling with longitudinal-optical (LO) phonons, where the temperature dependence of the scattering strength comes mainly from phonon occupation number \cite{39} $N_0=[e^{\hbar\omega_{LO}/k_BT}-1]^{-1}$ with $\hbar\omega_{LO}$ being the LO-phonon energy. For BL MoS$_2$, $\hbar\omega_{LO}\approx 49$ meV, which is obtained from the ``a'' peak in the Raman spectrum in Fig. \ref{fig2}(b) and is the same as that obtained from other studies \cite{40}; and iv) $1/\tau_I$, $\alpha_{AC}$ and $\Gamma_{LO}$ are temperature-independent fitting parameters, regarding to different scattering centers. By fitting (see the red section of Fig. \ref{fig5}), we obtain $\tau_I$=198.76 fs, $\alpha_{AC} =2.61 \times 10^{-2}$ fs$^{-1}$K$^{-1}$, and $\Gamma_{LO}=8.07 \times 10^{-6}$ fs$^{-1}$.

\subsection{Further discussions}
Now we discuss the important and unique features of 2$H$-stacking BL MoS$_2$ from a view point of physics. From the results obtained from our DFT calculation, as shown in Fig. \ref{fig2}(d), we notice the following points. i) In sharp contrast to ML MoS$_2$ in which the minima of the conduction band is at the $K$-point, the conduction band minima in 2$H$-stacking BL MoS$_2$ is at $\Sigma_{min}$-point \cite{36,25} between the $K$- and $\Gamma$-point. An indirect band gap can be found around the $\Sigma_{min}$-point. We find that the energy difference between the $K$-point and $\Sigma_{min}$-point in the conduction band is about 197 meV. As a result, the electrons occupy mainly the states around the $\Sigma_{min}$-point in a n-type 2$H$-stacking BL MoS$_2$. ii) In 2$H$-stacking BL MoS$_2$, the top of the valence band at the $\Gamma$-point is markedly higher than that at the $K$-point, which also differs sharply from ML MoS$_2$ in which the valence band maxima is at the $K$-point. Thus, the holes occupy mainly the states around the $\Gamma$-point in a p-type 2$H$-stacking BL MoS$_2$. However, an indirect band gap is presented around the $\Gamma$-point in 2$H$-stacking BL MoS$_2$. And iii) the features of the direct band gap around the $K$-point is relatively weak for 2$H$-stacking MoS$_2$.  These characteristics are the main reasons why the photon-induced light generation from 2$H$-stacking BL MoS$_2$ is significantly weaker than that from ML MoS$_2$, due to the presence of electronic transition channels accompanied by photon emission via indirect electron-photon coupling. As for PL emission from 2$H$-stacking BL MoS$_2$ via excitonic effect, it can only occur around the $K$-point because the direct electron-photon interaction normally does not vary the electronic momentum \cite{42,43}. Therefore, the inter-valley electronic transition is greatly impossible. Furthermore, for 2$H$-stacking BL MoS$_2$, two conduction bands around the $K$-point are roughly degenerated, whereas the valance bands are split. This is the reason why two PL peaks can be observed for 2$H$-stacking BL MoS$_2$ as shown in Fig. \ref{fig2}(c). It is shown that the splitting of the valence band also occurs in ML MoS$_2$ \cite{24} and in 3$R$-stacking BL MoS$_2$\cite{24}. Thus, the $B$-peak marked in Fig. \ref{fig2}(c) can be observed not only in 2$H$-stacking but also in ML and 3$R$-stacking BL MoS$_2$\cite{24}. However, the characteristic Raman peak ``a'' can help us to judge whether the BL structure is the 2$H$-stacking or the 3$R$-stacking \cite{14}.

After analyzing the theoretical results of the electronic band structure and the experimental results of PL measurement, we believe that the theoretical results obtained from the DFT for energy band of 2$H$-stacking BL MoS$_2$ are credible. Based on the results of the electronic band structure, we can also evaluate the effective electron mass $m^*$ in the conduction band around the $\Sigma_{min}$-point in 2$H$-stacking BL MoS$_2$, by fitting the band energy obtained from the DFT calculation with the energy spectrum $\hbar^2 k^2/(2m^*)$. In this work, we use $m^*=0.55 m_e$ \cite{36} to analysis the experimental results obtained from THz TDS measurement.

In a n-type 2$H$-stacking BL MoS$_2$, the conducting electrons occupy mainly the states around the $\Sigma_{min}$-point. Thus, in contrast to the PL emission, the electronic transition induced by THz irradiation is achieved mainly via intra-band channels around the $\Sigma_{min}$-point. The conduction band in low-energy regime around the $\Sigma_{min}$-point, shown in Fig. \ref{fig2}(d) for 2$H$-stacking BL MoS$_2$, is very much parabolic. As a result, the THz optoelectronic properties of a n-type 2$H$-stacking BL MoS$_2$ are semiconductor-like so that the theory of the optical conductivity for free-electrons can be applied to describe the features of the THz response in n-type 2$H$-stacking BL MoS$_2$. This is why we can use Eq. (\ref{2}) for a semiconductor to fit the temperature dependence of electronic relaxation time $\tau$ in BL MoS$_2$. We note that $\tau$ in 2$H$-stacking BL MoS$_2$ on sapphire substrate does not differ largely from that in ML MoS$_2$ on the same substrate \cite{16}. Similar to conventional semiconductors, $\tau$ in 2$H$-stacking BL MoS$_2$ decreases with increasing temperature (see the red section of Fig. \ref{fig5}). This is owing to the fact that the strength of electron interaction with acoustic-phonons increases and the strength of electron$-$LO-phonon coupling increases with temperature because the phonon occupation number $N_0$ increases with temperature.

We find that the electron density $n_e$ in a n-type 2$H$-stacking BL MoS$_2$ on sapphire substrate (see the blue section of Fig. \ref{fig5}) is much higher than that in ML MoS$_2$ \cite{16}. This is mainly because of a bilayer structure with inter-layer vdW interaction and the interaction between MoS$_2$ layers with the substrate. In comparison to ML MoS$_2$ on sapphire substrate, where $n_e$ increases significantly with increasing temperature, $n_e$ in 2$H$-stacking BL MoS$_2$ depends weakly on temperature. This implies that the donors in the BL-MoS$_2$/substrate system are most likely all ionized so that the effect of thermal ionization of the donors in the system becomes weak. The weak temperature dependence of $n_e$ in BL MoS$_2$ has also been observed through electronic transport measurements \cite{37}.

It should be noted that in temperature regime from 80 to 300 K, the $K$-point in conduction band of 2$H$-stacking BL MoS$_2$ may become occupied, so that two conducting channels from two valleys may contribute to THz conductivity. However, since the energy difference between the $K$-point and the $\Sigma_{min}$-point is about 197 meV obtained from our DFT calculation, the conducting electrons are largely populated in the $\Sigma_{min}$-band in 2$H$-stacking BL MoS$_2$. Furthermore, the effective electron mass in $\Sigma_{min}$-point is heavier than that in $K$-point \cite{44,45}. The electronic density-of-states (DoS) in the $\Sigma_{min}$-band are therefore larger than those in the $K$-band because the DoS is proportional to $m^*$ in a 2D electron gas system \cite{25}. Consequently, the contribution to $\sigma(\omega)$ from electronic transition in the $K$-band is very small so that we can employ the theoretical model for single transition channel for the analysis and fitting of the experimental results.

Moreover, we found previously that the presence of the dielectric substrate can induce the electronic backscattering or localization effect in ML TMD systems \cite{16,46}. As shown in Fig. \ref{fig4}(b), this effect can also be observed in BL MoS$_2$, where the complex optical conductivity satisfies largely the Drude-Smith formula. When BL MoS$_2$ is placed on the substrate, charged impurities and phonons in the substrate can provide the additional scattering centers for electrons in BL MoS$_2$. This effect has been investigated in the graphene-based device \cite{47,48}. These additional scattering centers can further randomize the time dependence of the momentum and energy distribution of the electrons in BL MoS$_2$. Thus, the effect of electronic backscattering or localization can be enhanced by the presence of the substrate. Because the scattering rate increases with temperature, the effect of substrate-induced electronic localization also increases in BL MoS$_2$ on sapphire substrate (see inset in Fig. \ref{fig5}), noting that $c$ is always a negative value.

\section{Conclusions}
In this study, we have fabricated the 2$H$-stacking BL MoS$_2$ on sapphire substrate using the standard CVD growth and PDMS transfer techniques. The THz TDS has been applied to measure the optoelectronic properties of this kind of TMD based BL system. The theoretical calculation on the basis of the DFT approach has been carried out for understanding of the experimental findings. The main conclusions drawn from this work can be summarized as follows.

The electronic band structure in 2$H$-stacking BL MoS$_2$ differs markedly from that in ML MoS$_2$. As a result, the weak PL emission from 2$H$-stacking BL MoS$_2$ is mainly induced by exciton effect occurring around the $K$-point. In contrast, the THz response is mainly achieved through intra-band electronic transition accompanied by the absorption of photons around the $\Sigma_{min}$-point which is in-between the $K$- and $\Gamma$-point in 2$H$-stacking BL MoS$_2$.

Through THz TDS measurement, we can obtain the complex optical conductivity for 2$H$-stacking BL MoS$_2$. It is found that $\sigma(\omega)$ for 2$H$-stacking BL MoS$_2$ sapphire substrate does not obey the conventional Drude formula for optical conductivity. But it can be reproduced by the Drude-Smith formula in which the effect of photon-induced electronic backscattering has been considered. By fitting the experimental results with the theoretical formula, we are able to determine optically the key sample parameters in BL MoS$_2$, such as the electron density, the electronic relaxation time, and the electronic localization factor. The temperature dependence of these parameters has been examined. We have found that the THz optoelectronic properties of 2$H$-stacking BL MoS$_2$ are semiconductor-like, where the electron density depends weakly on temperature in 80 K to 300 K regime and the electronic relaxation time decreases with increasing temperature owing to the presence of electron-phonon scattering. Furthermore, similar to ML MoS$_2$ on a dielectric substrate, the substrate-induced electronic backscattering or localization can also be observed for BL MoS$_2$ on sapphire substrate. It should be noticed that by using THz TDS, we do not need the external magnetic field for the determination of the electron density via, e.g., the Hall measurement.

2$H$-stacking BL MoS$_2$ is a basic structure in twistronic system. The other five high-symmetric stacking configurations can be achieved by twisting and/or translating one of the MoS$_2$ layer away from 2$H$-stacking. Therefore, the investigation of 2$H$-stacking BL MoS$_2$ is of great importance and significance for the understanding of TMD based BL twistronic systems. We hope that the results obtained from this study can contribute an in-depth understanding of the basic optoelectronic properties of bilayer TMD materials and can be the basis for the application of BL MoS$_2$ in advanced optical and optoelectronic devices.

\section{Conflicts of interest}
There are no conflicts to declare.

\section*{ACKNOWLEDGMENTS}
This work was supported by the National Natural Science foundation of China (NSFC) (Grants No. U2230122 and No. U2067207)
and by Shenzhen Science and Technology Program (No. KQTD20190929173954826).The numerical calculations in this work were conducted at Hefei advanced computing center.


\begin{thebibliography}{99}

\bibitem{1}
K.~S. Novoselov, A.~K. Geim, S.~V. Morozov, D.~Jiang, Y.~Zhang, S.~V. Dubonos,
  I.~V. Grigorieva, and A.~A. Firsov,
{Science, 2004, \textbf{306}, 666-669}.

\bibitem{2}
C.~Castellani, C.~DiCastro, P.~A. Lee,
{Phys. Rev. B, 1998, \textbf{57}, R9381-R9384}.

\bibitem{3}
M.~K. Fai, X.~Di, S.~Jie,
{Nat. Photonics, 2018, \textbf{12}, 451-460}.

\bibitem{4}
H.~M. Hill, A.~F. Rigosi, C.~Roquelet, A.~Chernikov, T.~F. Heinz,
{Nano Lett., 2015, \textbf{15}, 2992-2997}.

\bibitem{5}
D.~Xiao, G.~B. Liu, W.~Feng, X.~Xu, W.~Yao,
{Phys. Rev. Lett., 2012, \textbf{108}, 196802}.

\bibitem{6}
Q.~J. Tong, H.~Yu, Q.~Zhu, Y.~Wang, X.~Xu, Y.~Wang,
{Nat. Phys., 2017, \textbf{13}, 356-362}.

\bibitem{7}
Y.~Cao, V.~Fatemi, S.~Fang, K.~Watanabe, T.~Taniguchi, E.~Kaxiras, P.~Jarillo-Herrero,
{Nature, 2018, \textbf{556}, 43-50}.

\bibitem{8}
K.~Seyler, P.~Rivera, H.~Yu, N.~Wilson, E.~Ray, D.~Mandrus, J.~Yan, W.~Yao, X.~Xu,
{Nature, 2019, \textbf{567}, 66-70}.

\bibitem{9}
T.~Cai, S.~A. Yang, X.~Li, F.~Zhang, J.~Shi, W.~Yao, Q.~Niu,
{Phys. Rev. B, 2013, \textbf{88} 115140}.

\bibitem{10}
K.~F. Mak, C.~Lee, J.~Hone, J.~Shan, T.~F. Heinz,
{Phys. Rev. Lett., 2010, \textbf{105} 136805}.

\bibitem{11}
C.~G. Lee, H.~G. Yan, E.~B. Louis, F.~H. Tony, H.~James, R.~Sunmin,
{{ACS} Nano, 2010, \textbf{4}, 2695-2700}.

\bibitem{12}
J.~E. Padilha, H.~Peelaers, A.~Janotti, C.~G. Van, de~Walle,
{Phys. Rev. B, 2014, \textbf{90}, 205420}.

\bibitem{13}
X.~Z. Zhang, R.~Y. Zhang, Y.~Zhang, T.~Jiang, C.~Y. Deng, X.~A. Zhang, S.~Q. Qin,
{Opt. Mater., 2019, \textbf{94}, 213-216}.

\bibitem{14}
M.~Xia, B.~Li, K.~Yin, G.~Capellini, Y.~H. Xie,
{{ACS} Nano, 2015, \textbf{9}, 12246-12254}.

\bibitem{15}
A.~M. Van, J.~Kunstmann, A.~Chernikov, D.~A. Chenet, Y.~M. You, X.~X. Zhang, P.~Y. Huang, T.~C. Berkelbach, L.~Wang, F.~Zhang,
{Nano Lett., 2014, \textbf{14}, 3869-3875}.

\bibitem{16}
C.~Wang, W.~Xu, H.~Mei, H.~Qin, X.~Zhao, C.~Zhang, H.~F. Yuan, J.~Zhang, Y.~Xu, P.~Li, M.~Li,
{Opt. Lett., 2019, \textbf{44}, 2139-2142}.

\bibitem{17}
H.~Wen, W.~Xu, C.~Wang, D.~Song, H.~Y. Mei, J.~Zhang, L.~Ding,
{Nano Select, 2020, \textbf{2} 90-92}.

\bibitem{18}
M.~Liao, Z.~Wei, L.~Du, Q.~Q. Wang, T.~Jiang, H.~Yu, F.~Wu, J.~J. Zhao, X.~Xu, B.~Han,
{Nat. Commun., 2020, \textbf{11}, 2153}.

\bibitem{19}
H.~Yu, M.~Z. Liao, W.~J. Zhao, G.~D. Liu, G.~Y. Zhang,
{ACS Nano, 2017, \textbf{11}, 12001-12007}.

\bibitem{20}
Y.~Yu, C.~Li, Y.~Liu, L.~Su, Y.~Zhang, L.~Cao,
{Sci. Rep., 2013, \textbf{3}, 1866}.

\bibitem{21}
X.~Wang, H.~Feng, Y.~Wu, L.~Jiao,
{J. Am. Chem. Soc., 2013, \textbf{135}, 5304-5307}.

\bibitem{22}
H.~Sajjad, A.~S. Muhmmad, V.~Dhanasekaran, Z.~I. Muhmmad, S.~Jai, F.~K. Muhmmad, E.~Jonghwa, S.~Yongho, J.~Jongwan,
{J. Alloy. Compd., 2015, \textbf{653} 369-378}.

\bibitem{23}
C.~R. Zhu, G.~Wang, B.~L. Liu, X.~Marie, X.~F. Qiao,
{Phys. Rev. B., 2013, \textbf{88}, 121301}.

\bibitem{24}
F.~Ullah, J.~H. Lee, Z.~Tahir, A.~Samad, C.~T. Le, J.~Kim, D.~Kim, M.~U. Rashid, S.~Lee, K.~Kim, H.~Cheong, J.~I. Jang, M.~J. Seong, Y.~S. Kim,
{{ACS} Appl. Mater. $\&$Inter., 2021, \textbf{13}, 57588-57596}.

\bibitem{25}
J.~K. Ellis, M.~J. Lucero, G.~E. Scuseria,
{Appl. Phys. Lett., 2011, \textbf{99}, 261908}.

\bibitem{26}
S.~Bhattacharyya, A.~K. Singh,
{Phys. Rev. B, 2012, \textbf{86}, 075454}.

\bibitem{27}
P.~L. Christiansen, M.~P. Srensen, A.~C. Scott,
{Nonlinear Science at the Dawn of the 21st Century. (Springer Berlin Heidelberg, 2000)}.

\bibitem{28}
M.~Hangyo, T.~Nagashima, S.~Nashima,
{Meas. Sci. Technol., 2002, \textbf{13}, 1727}.

\bibitem{29}
L.~Duvillaret, F.~Garet, J.-L. Coutaz,
{IEEE J. Sel. Top. Quant., 1996, \textbf{2}, 739-746}.

\bibitem{30}
S.~Nudelman, S.~S. Mitra,
{Optical Properties of Solids. (Springer, 1969)}.

\bibitem{31}
M.~Tinkham,
{Phys. Rev. B, 1956, \textbf{104}, 845}.

\bibitem{32}
J.~D. Jackson,
{Classical Electrodynamics, 3rd Edition. (SWiley, 1998)}.

\bibitem{33}
P.~Drude,
{Bestimmung der optischen Constanten der Metalle. (Annalen Der Physik, 1890)}.

\bibitem{34}
N.~V. Smith,
{Phys. Rev. B, 2001 \textbf{64}, 155106}.

\bibitem{35}
F.~W. Han, W.~Xu, L.~L. Li, C.~Zhang,
{J. Appl. Phys., 2016, \textbf{119}, 245706}.

\bibitem{36}
A.~Mukhopadhyay, S.~Kanungo, H.~Rahaman,
{J. Computat. Electron., 2021, \textbf{20}, 161-168}.

\bibitem{37}
B.~Baugher, H.~Churchill, Y.~Yang, P.~Jarillo-Herrero,
{Nano Lett., 2013, \textbf{13}, 4212-4216}.

\bibitem{38}
D.~Valerini, A.~Cret\'{\i}, M.~Lomascolo, L.~Manna, R.~Cingolani, M.~Anni,
{Phys. Rev. B, 2005, \textbf{71}, 235409}.

\bibitem{39}
W.~Xu, F.~M. Peeters, T.~C. Lu,
{Phys. Rev. B, 2009, \textbf{79}, 037403}.

\bibitem{40}
S.~Kim, A.~Konar, W.~Hwang, L.~Jong-hak, J.~Lee, J.~Yang, C.~Jung, H.~Kim, J.~W. Yoo, J.~H. Choi, Y.~Jin, S.~Lee, D.~Jena, W.~Choi, K.~Kim,
{Nat. commun., 2012, \textbf{3}, 1011}.

\bibitem{42}
H.~Schwarz,
{Laser Interaction and Related Plasma Phenomena. (Springer US, 1972)}.

\bibitem{43}
W.~Xu, H.~M. Dong, L.~L. Li, J.~Q. Yao, P.~Vasilopoulos, F.~M. Peeters,
{Phys. Rev. B, 2010, \textbf{82}, 125304}.

\bibitem{44}
W.~S. Yun, S.~W. Han, S.~C. Hong, I.~G. Kim, J.~D. Lee,
{Phys. Rev. B, 2012, \textbf{85}, 033305}.

\bibitem{45}
E.~Scalise, M.~Houssa, G.~Pourtois, V.~Afanas, A.~Stesmans,
{Physica E, (2014), \textbf{56}, 416-421}.

\bibitem{46}
H.~M. Dong, Z.~H. Tao, L.~L. Li, F.~Huang, W.~Xu, F.~M. Peeters,
{Appl. Phys. Lett., 2020 \textbf{116}, 203108}.

\bibitem{47}
H.~M. Dong, W.~Xu, Z.~Zeng, T.~C. Lu, F.~M. Peeters,
{Phys. Rev. B, 2008, \textbf{77}, 235402}.

\bibitem{48}
S.~H. Zhang, W.~Xu, S.~M. Badalyan, F.~M. Peeters,
{Phys. Rev. B, 2013, \textbf{87} 075443}.

\bibitem{49}
S. Kumar, A. Singh, S. Kumar, A. Nivedan, M. Tondusson, J. Degert, J. Oberle, S. J. Yun, Y. H. Lee and E. Freysz,
{Opt. Express, 2021, \textbf{87}, 4181-4190}.

\bibitem{50}
M. Bala Murali Krishna, J. Mad\'eo, J. P. Urquizo, X. Zhu, C. Vinod S Shekar Tiwary, M. P. Ajayan and M. K. Dani,
{Semicond. Sci. Technol., 2018, \textbf{33}, 084001}.


\end{thebibliography}
\end{document}